\documentstyle[12pt]{article}

\catcode`\@=11
\def\marginnote#1{}
\newcount\hour
\newcount\minute
\newtoks\amorpm
\hour=\time\divide\hour by60
\minute=\time{\multiply\hour by60 \global\advance\minute by-\hour}
\edef\standardtime{{\ifnum\hour<12 \global\amorpm={am}%
        \else\global\amorpm={pm}\advance\hour by-12 \fi
        \ifnum\hour=0 \hour=12 \fi
        \number\hour:\ifnum\minute<10 0\fi\number\minute\the\amorpm}}
\edef\militarytime{\number\hour:\ifnum\minute<10 0\fi\number\minute}

\def\draftlabel#1{{\@bsphack\if@filesw {\let\thepage\relax
   \xdef\@gtempa{\write\@auxout{\string
      \newlabel{#1}{{\@currentlabel}{\thepage}}}}}\@gtempa
   \if@nobreak \ifvmode\nobreak\fi\fi\fi\@esphack}
        \gdef\@eqnlabel{#1}}
\def\@eqnlabel{}
\def\@vacuum{}
\def\draftmarginnote#1{\marginpar{\raggedright\scriptsize\tt#1}}
\def\draft{\oddsidemargin -.5truein
        \def\@oddfoot{\sl preliminary draft \hfil
        \rm\thepage\hfil\sl\today\quad\militarytime}
        \let\@evenfoot\@oddfoot \overfullrule 3pt
        \let\label=\draftlabel
        \let\marginnote=\draftmarginnote
   \def\@eqnnum{(\theequation)\rlap{\kern\marginparsep\tt\@eqnlabel}%
\global\let\@eqnlabel\@vacuum}  }

\def\numberbysection{\@addtoreset{equation}{section}
        \def\theequation{\thesection.\arabic{equation}}}

\def\underline#1{\relax\ifmmode\@@underline#1\else
        $\@@underline{\hbox{#1}}$\relax\fi}

\catcode`@=12
\relax

\numberbysection
\begin{document}

\begin{titlepage}
\nopagebreak
\begin{flushright}

hep-th/9211088
\\
DIAS--92/27
\\
IHEP 92-155
\\
LPTENS--92/40
\\
NI92008
\\
November 1992
\end{flushright}

\vglue 2  true cm
\begin{center}
{\large\bf GAUGE CONDITIONS  FOR THE \\
CONSTRAINED-WZNW--TODA REDUCTIONS}
\vglue 1  true cm
{\bf Jean-Loup~GERVAIS}\\
{\footnotesize
Laboratoire de Physique Th\'eorique de l'\'Ecole Normale
Sup\'erieure\footnote{Unit\'e Propre du Centre National de la
Recherche Scientifique, associ\'ee \`a l'\'Ecole Normale
Sup\'erieure et \`a l'Universit\'e de Paris-Sud.},\\ 24 rue Lhomond,
75231 Paris C\'EDEX 05, ~France.}\\
{\bf Lochlainn~O'RAIFEARTAIGH}\\
{\footnotesize School of Theoretical Physics, Dublin Institute for Advanced
Studies,\\ 10 Burlington  Road, Dublin 4, ~Ireland.}\\
{\bf Alexander V.~RAZUMOV}\\
{\footnotesize  Institute for High Energy Physics,\\ 142284, Protvino, Moscow
region, ~Russia. \\and\\ }
{\bf Mikhail V.~SAVELIEV}
\footnote{ On leave of absence from the Institute for High
Energy Physics, 142284, Protvino, Moscow region, ~Russia.}\\
{\footnotesize Isaac Newton Institute for Mathematical Sciences, University
of Cambridge,\\ 20 Clarkson Road, Cambridge CB3 0EH, ~U.K.}\\
\medskip
\end{center}

\vfill
\begin{abstract}
\baselineskip .4 true cm
\noindent

{\footnotesize There is a constrained-WZNW--Toda
theory for any simple
Lie algebra equipped with an  integral gradation.
It is explained  how the different approaches to
these dynamical systems  are related by gauge
transformations.
 Combining  Gauss decompositions in
 relevent  gauges, we unify formulae
already derived, and
explictly  determine
the holomorphic expansion of the
conformally reduced WZNW solutions ---
whose restriction gives the  solutions  of the
Toda equations. The same takes place
also for semi-integral  gradation. Most of our conclusions
are also applicable to
the affine Toda theories. }

\end{abstract}
\vfill
\end{titlepage}
\baselineskip .5 true cm

\section{Introduction}
Two-dimensional conformal Toda systems have
 been invented and explicitly integrated
on the basis  of the group-algebraic approach more than ten years ago
in papers\footnote{In
what follows, and when we use
 the results obtained by the authors of refs.\cite{1}, \cite{2},
in the framework of the algebraic approach, we give
reference to their review book\cite{3}
(of course,
if the result is described
there) rather that to
the  original papers\cite{1}, \cite{2}.}
 \cite{1} and \cite{2}.
They are in general associated with any integral
gradation of a simple Lie algebra.  We shall deal with the general
case, that is, not only with the principal
gradation that leads to the
 so-called
 finite non-periodic Toda systems,
but also with their
non-abelian
versions\footnote{Here we deal with  systems that are different from
those proposed by A. M. Polyakov
at the end of the seventies, and also
called by him  ``nonabelian Toda theories''.} introduced
in refs.\cite{2}.
 In what follows, and  for the
sake of brevity, we call them all Toda-type systems.
Some time after refs.\cite{1}\cite{2},
a number of investigations appeared  where these systems
arose in the context of two dimensional gravity,
of various aspects of
W-algebras and W-geometries,  of constrained  WZNW-models,
of   ``topological and
anti-topological fusion'', and so on,
-- see \cite{G}-\cite{11} and references
therein. At present, however,  the link between these  papers is
not easy to make, although they basically deal with the
same objects and notions. Different notations have
often been used, and, more importantly, equivalent formulae
may be connected only when one sees that
 for the various problems it was convenient
to use different gauges.

This disaccordance
 is quite unsuitable to make further progress.
This is true, not only for
concrete applications, but also
for the possibility (and desirability) of
extending the ideas to multi-dimensional systems following a
path opened, e.g.  in refs.\cite{9}, \cite{11}.
Accordingly, the present paper is partly a unifying review
of   the various
gauge conditions that have been used and of the relationship between
them. Our contribution  is, to some extent,
 a methodological one
proposed for a wide audience of people interested in
varous aspects of  integrable
systems and related subjects. Systematically combining Gauss
decompositions in these different gauges will nevertheless
lead us to novel results, such as the explicit formula
for the holomorphic decomposition of the WZNW solution.

\bigskip

\section{The basic framework}

We consider two--dimensional space with coordinates $z^+$, $z^-$.
In this discussion, the unifying feature is the zero-curvature
condition -- also called Maurer-Cartan, Lax, and so on:
\begin{equation}
[\partial _+ + A_+, \partial _- + A_-]_- \equiv \partial _+A_--
\partial _-A_++[A_+, A_-]_-=0.
\label{2}
\end{equation}
As usual, $\partial_\pm$ stands for $\partial /\partial z^\pm$.
$A_\pm$ takes values in a simple finite-dimensional
Lie algebra ${\cal G}$. One considers a $Z$-gradation
\[
{\cal G}=\oplus _{m\in {\bf Z}}{\cal G}_m \equiv {\cal G}_-\oplus
{\cal G}_0 \oplus {\cal G}_+,
\]
and, associated with this choice, we have the modified Gauss
decompositions, such that
any regular element $U$  of the corresponding
Lie group $G$ may be written under the form\footnote{Of course
there are equivalent
decompositions such as $U= U'_{(-)} U'_{(0)} U'_{(+)} $ where the
elements of $ G_\pm$
and $ G_0$ are in different orders.}
\begin{equation}
U= U_{(+)} U_{(0)} U_{(-)},\>  U_{(\pm)}\in G_\pm,\>
U_{(0)}\in  G_0.
\label{gd}
\end{equation}
$G_\pm$ are the nilpotent algebras generated by
${\cal G}_\pm$, and $G_0$ is the subgroup  generated by ${\cal G}_0$.
Note that ${\cal G}_0$ may be a non-abelian algebra   -- just
then the Toda
theory is  called non-abelian. In Eq.\ref{gd} we introduced a
notation  which we will use throughout the article:
group elements that belong to
$G_\pm$ or  $ G_m$ are given    the same
subscripts in parenthesis. We shall also use a similar
notational rule for elements of ${\cal G}_\pm$ or  $ {\cal G}_m$

Of course
the zero-curvature condition Eq.\ref{2} is invariant under
the general gauge transformation
\begin{equation}
A_\pm \to A^h_{\pm} = h^{-1} A_\pm \,h  + h^{-1}\partial_\pm h,
\label{gt}
\end{equation}
 where $h$ is any element of $G$.
A basic requirement of the Toda dynamics is that
the above vector potentials  satisfy the condition
$A_\pm \in {\cal G}_0 \oplus {\cal G}_{\pm }$ in some gauges.
These gauges will be called triangular.
This   triangular gauge condition  is left
invariant by the gauge transformations
\begin{equation}
A_\pm \to A^{h_{(0)}}_{\pm} =
h_{(0)}^{-1} A_\pm \,h_{(0)} +
h_{(0)}^{-1}\partial_\pm \,h_{(0)}, \>
\hbox{with} \> h_{(0)} \in G_0.
\label{66}
\end{equation}
 We shall call  them
$G_0$-gauge transformations, in the present paper.
The abelian and non-abelian Toda
theories  have  abelian and non-abelian $G_0$-gauge groups
respectively.
Obviously the solution of the zero-curvature condition is
\begin{equation}
A_{\pm}=g^{-1}\partial _{\pm}g,
\label{6b}
\end{equation}
with $g\in G$. Thus under the gauge transformation Eq.\ref{gt},
\begin{equation}
g\to g h.
\label{6g}
\end{equation}
Following ref.\cite{2},
let us recall that  the Toda equation may be written as\footnote{
Note also that
the equation $\partial _{+i}(\hat {g}_{(0)}^{-1}\partial _{-j}
\hat {g}_{(0)})=[X_{-j}, \hat {g}_{(0)}^{-1}X_{i}\hat {g}_{(0)}],
\,1\leq i,j\leq n$, recently proposed in \cite{12} as a natural
generalization of the so-called special geometry, represents, at least
formally, a multi-dimensional analogue of Eq.\ref{T}. This
generalized equation may probably be
 considered as an appropriate reduction of
the 2n-dimensional WZNW-model\cite{9},
just with the same reasonings as in \cite{4}
for the usual 2D case.}

\begin{equation}
\partial _+(\hat {g}_{(0)}^{-1}\partial _-\hat {g}_{(0)})=
[X_{(-1)}, \hat {g}_{(0)}^{-1}X_{(1)}\hat {g}_{(0)}].
\label{T}
\end{equation}
where $X_{(\pm 1)}$ are fixed elements of
${\cal G}_{\pm 1}$ such that
the whole ${\cal G}_{\pm 1}$   may be recovered by adjoint action of
${\cal G}_0$, and where the dynamical fields are lumped into
$\hat {g}_{(0)}(z_+,z_-)\in G_0$. We shall see how this is
connected with Eq.\ref{2}
 in different gauges. Let us mention   some well known examples.
The  elements $X_{(\pm 1)}$ could be the
generators of a $sl(2, {\bf
C})$-subalgebra of ${\cal G}$, and the
corresponding ${\bf Z}$-gradation be
performed by the Cartan element $H=[X_{(1)}, X_{(-1)}]$ of
this $sl(2, {\bf C})$, i.e.,
$[H, {\cal G}_m]=m\,{\cal G}_m$.
In particular, for the principal (canonical) gradation of a
simple Lie algebra ${\cal G}$ (with the abelian
subalgebra ${\cal G}_0$
and $X_{(\pm 1) }$ decomposable over the
Chevalley generators of the positive
and negative simple roots, respectively), equation (\ref{T}) reduces
to the abelian Toda system
\begin{equation}
\partial_+ \partial_-
x_i=\exp \rho_i, \;1\leq i \leq r \equiv \mbox{
rank }
{\cal G};\; \rho_i=\sum_{j=1}^rk_{ij}x_j,
\label{3}
\end{equation}
where $x_i(z_+,z_-)$ are the Toda
fields,  $k$ is the Cartan matrix of ${\cal G}$.
Recall here the defining
relations for the Cartan ($h_i$) and Chevalley
($X_{\pm  i}$) generators
of ${\cal G}$:
\begin{equation}
{}[h_i, h_j]=0,\;[h_i, X_{\pm  j}]=\pm k_{ji}X_{\pm j},\;
[X_{i}, X_{-j}]=\delta _{ij}h_i;\;1\leq i,j\leq r.
\label{4}
\end{equation}
Of course, the simple Lie algebras ${\cal G}$,
in general, could be also
supplied with non-principal gradations,
in particular those associated with
non-principal $sl(2, {\bf C})$
subalgebras of ${\cal G}$. For such gradations
the subalgebra ${\cal G}_0$ is not abelian.

The Toda field equations Eq.\ref{T}
follow from the zero-curvature condition
if one requires that there exist triangular gauges where
\begin{equation}
A_\pm \in {\cal G}_0 \oplus {\cal G}_{\pm 1}.
\label{gr}
\end{equation}
We shall characterize these gauges as being restricted-triangular.
Of course, since these conditions  are not invariant
under the general transformations Eq.\ref{gt}, we may only require
that there exist  gauges  such that the vector
potentials take certain  restricted forms.
It is clearly  not true that any vector potential
 may be gauge-transformed to become restricted-triangular.
Thus  the restricted-triangularity condition
 is not a matter of gauge choice.
In fact,  {\bf it determines the
dynamics}.
Other hypothesis may be contradictory, that is,
only give trivial
dynamics (such as $A_{\pm} \in {\cal G}_{\mp m}
\oplus {\cal G}_0 \oplus {\cal G}_{\pm 1}$); or lead to higher
Toda flows (such as $A_{\pm} \in {\cal G}_0 \oplus
{\cal G}_{\pm m}$,
 with $m>1$).
The equations associated with the last possibility also are integrable
\cite{3}. The  holomorphic expansions
in different gauges could  be studied,
following the method described here,
but we shall not do so, at present.

\section{Relevent  gauges}

In this section we discuss three different
gauges which are relevent for this
or that problems under consideration, namely ``general triangular'',
``Toda'', and
``WZNW'' gauges.

\subsection{General triangular gauges}

To begin with, we  discuss properties that hold in any
triangular gauge,
without assuming that the restriction condition Eq.\ref{gr} holds.
We distinguish these gauges  by a superscript $(t)$.
This is essentially a  review of some
 results from \cite{3}.
Two  useful  forms  for
 the    Gauss decomposition of the corresponding $g^{(t)}$
of Eq.\ref{6b} are as follows
\begin{equation}
g^{(t)}= M_{(-)}N_{(+)}g_{(0)-}=
M_{(+)}N_{(-)}g_{(0)+},
\label{7}
\end{equation}
Under the $G_0$-gauge transformation
Eq.\ref{66}, $M_{(\pm)}$,
$N_{(\pm)}$ are clearly
invariant.
 On the other hand,
$g^{(t)}_{(0)\pm} \to g^{(t)}_{(0)\pm} h_{(0)}$, and we introduce
the $G_0$-gauge-invariant.
\begin{equation}
g_{(0)}\equiv g_{(0)+}g^{ -1}_{(0)-}
\label{g0}
\end{equation}
Next, it is easily seen that the triangular gradation
properties  are satisfied iff  $M_{(\pm)}$
are holomorphic functions, that is
\begin{equation}
\partial _{\mp}M_{(\pm)}=0,
\label{h}
\end{equation}
and one deduces from Eq.\ref{6b} that\footnote{For simplicity of
notation, $\partial_\pm$ only act on the first following function,
unless parenthesis indicate otherwize. }
\begin{eqnarray}
A^{(t)}_+ & = & g^{ -1}_{(0)-}
N_{(+)}^{-1}\partial _+N_{(+)}g_{(0)-}+
g^{ -1 }_{(0)+}\partial _+g_{(0)-},\nonumber\\
A^{(t)}_- & = &
g^{ -1}_{(0)+}
N_{(-)}^{-1}\partial _-N_{(-)}g_{(0)+}+
g^{ -1 }_{(0)+}\partial _-g_{(0)+}.
\label{8}
\end{eqnarray}
According to Eq.\ref{h},
 one may define two holomorphic functions $L_{(\pm)}(z_\pm)
\in {\cal G}_{\pm }$ such that
\begin{equation}
dM_{(\pm)} \bigl / dz_\pm = L_{(\pm )} M_{(\pm)}.
\label{L}
\end{equation}
On the other hand, it follows from Eqs.\ref{7}, and \ref{g0} that
\begin{equation}
K\equiv M_{(+)}^{-1}M_{(-)}=N_{(-)}g_{(0)}N_{(+)}^{-1}.
\label{10}
\end{equation}
The differential equations \ref{L} give
\begin{equation}
K^{-1}\, \partial _-K=L_{(-)}, \; \partial _+K\,  K^{-1}=-L_{(+)}.
\label{11}
\end{equation}
Substituting in these equations the element $K$ expressed via
the second part of the formula (\ref{10}), we arrive at
\begin{equation}
N_{(-)}^{-1}\partial _-N_{(-)}=
g_{(0)} L_{(-)}g_{(0)}^{-1},\quad  N_{(+)}^{-1}\partial _+N_{(+)}=
g^{-1}_{(0)} L_{(+)}g_{(0)}.
\label{12}
\end{equation}
It finally follows from Eqs.\ref{g0}, \ref{8},
and \ref{12} that, in any triangular gauge,
\begin{eqnarray}
A^{(t)}_+=g^{ -1}_{(0)+}
L_{(+)}g_{(0)+}+g^{ -1 }_{(0)-}
\partial _+g_{(0)-},
\nonumber\\
A^{(t)}_-=g^{ -1}_{(0)-}L_{(-)}
g_{(0)-}+g^{ -1}_{(0)+}\partial _
-g_{(0)+}.
\label{13}
\end{eqnarray}

\subsection{Toda gauges}
For the Toda dynamics, there exist   restricted-triangular
gauges where
the more restrictive condition Eq.\ref{gr} holds. These gauges
will be characterized by a superscript $(rt)$. If one is
in such a gauge, and
using the formulae just written, it is easy to see that,
$N_{(\pm)} \in  G_{\pm 1} $, $M_{(\pm)}\in    G_{\pm 1} $,
and $L_{(\pm)} \in {\cal G}_{\pm 1}$. In accordance with our general
convention, we may write them as $N_{(\pm 1 )}$, $M_{(\pm 1)}$,
$L_{(\pm 1)}$. According to ref.\cite{3},
 $L_{(\pm 1)}$
may be obtained from fixed elements $X_{(\pm 1)} \in {\cal G}_{\pm 1}$
by adjoint action of $G_0$. Thus we may write
\begin{equation}
L_{(\pm 1)} = y_{(0)\pm} X_{(\pm 1)} y^{-1}_{(0)\pm},
\label{XX}
\end{equation}
where, according to Eqs.\ref{h}, \ref{L},
$y_{(0)\pm}$ are holomorphic:
$\partial _\pm y_{(0)\mp}=0$.
Therefore
\begin{eqnarray}
A^{(rt)}_+=\left ( y^{-1}_{(0)+}\, g_{(0)+}\right )^{-1}X_{(1)}
\left ( y^{-1}_{(0)+}\, g_{(0)+}\right )
+g^{ -1}_{(0)-}\partial _+g_{(0)-},
\nonumber\\
A^{(rt)}_-=\left ( y^{-1}_{(0)-}\, g_{(0)-}\right )^{-1}X_{(-1)}
\left ( y^{-1}_{(0)-}\, g_{(0)-}\right )
+g^{ -1 }_{(0)+}\partial _-g_{(0)+}.
\label{AA}
\end{eqnarray}
The precise form Eq.\ref{T} of the Toda equation comes out
directly from the zero-curvature condition with specific
conditions introduced
in ref.\cite{3}. There are two associated gauge choices which
we call Toda gauges, and distinguish with a superscript $(T)$.
They may be characterized by the fact that
the gradation-zero  part in one of the connections, say $A^{(T)}_+$
is set equal
to zero, while the gradation-one component of the other
$A^{(T)}_-$ is  chosen to be constant. Next we verify  that,
  starting from any restricted-triangular gauge, there
exists  a $G_0$-gauge transformation
\begin{equation}
A_{\pm}^{(T)}\equiv h_{(0)}^{-1}(A^{(rt)}_{\pm}
+\partial _{\pm})h_{(0)},\;
\mbox{ i.e.,}\; g^{(T)}\equiv g^{(rt)}h_{(0)},
\label{14}
\end{equation}
such that
\begin{eqnarray}
A_{(0)+}^{(T)}&=& \left (g_{(0)-}\, h_{(0)}\right )^{-1}
\partial_ +\left (g_{(0)-}\, h_{(0)}\right ) =0, \nonumber \\
A_{(1)-}^{(T)}&=
&\left ( y^{-1}_{(0)-}\, g_{(0)-}\, h_{(0)}\right )^{-1}X_{(-1)}
\left ( y^{-1}_{(0)-}\, g_{(0)-}\,  h_{(0)}\right )
=\>\hbox{constant}.
\label{gg}
\end{eqnarray}
In the last equations we made use of Eqs.\ref{AA}.
Indeed, it is enough to choose
\begin{equation}
h_{(0)}= g_{(0)-}^{-1}\,  y_{(0)-}.
\label{ch}
\end{equation}
Next,  it is easily seen  that the Toda equations \ref{T}
 come out from the
flatness condition Eq.\ref{2} in this gauge, if we identify
\begin{equation}
\hat {g}_{(0)}=y_{(0)+}^{-1}\, g_{(0)+}\, h_{(0)},
\label{ghat}
\end{equation}
and one gets
\begin{equation}
A_{+}^{(T)}=\hat {g}_{(0)}^{-1}X_{(1)}\hat {g}_{(0)},
\;A_{-}^{(T)}=\hat {g}_{(0)}^{-1}\partial _-\hat {g}_{(0)}+X_{(-1)}.
\label{15}
\end{equation}
Finally, one may rewrite $h_{(0)}$, $g^{(T)}$, and $\hat g_0$
in terms of the original quantities introduced above for
an arbitrary restricted triangular gauge, obtaining,
\begin{eqnarray}
h_{(0)}\, & = & (y_{(0)-}\, g_{(0)-})^{-1}=g_{(0)+}^{-1}\,
y_{(0)+}\hat {g}_{(0)};\\
\label{16}
g^{(T)} & = & g^{(rt)}\, g_{(0)-}^{-1}\,y^{-1}_{(0)-};\\
\label{17}
\hat {g}_{(0)}\, & = &y_{(0)+}^{-1}
\, g_{(0)+}\, g_{(0)-}^{-1}\, y_{(0)-}\equiv
y_{(0)+}^{-1}
\, g_{(0)}\, y_{(0)-}.
\label{18}
\end{eqnarray}
Note that the functions (\ref{15})
for the case of the abelian Toda system
({\ref{3}) are reduced to the form
\begin{equation}
A_{+}^{(T)}=\sum _j e^{\rho _j}\,X_{j},\;
A_{-}^{(T)}=-\sum _j\partial _- x_j\,h_j+\sum _j X_{-j}.
\label{20}
\end{equation}
where the sum runs over the step operators associated with
simple    roots.

The role of the components $A_+$ and $A_-$ in the Toda gauge
may be, of course,
reversed by letting
\begin{equation}
\tilde {A}_{\pm }^{(T)}=\hat {g}_{(0)} A_{\pm }^{(T)}
\hat {g}_{(0)}^{-1} +
\hat {g}_{(0)} \partial _{\pm }\hat {g}_{(0)}^{-1},
\end{equation}
so that
\begin{equation}
\tilde {A}_{+}^{(T)}=\hat {g}_{(0)}\partial _+
\hat {g}_{(0)}^{-1}+X_{(1)},\;
\tilde {A}_{-}^{(T)}=\hat {g}_{(0)}X_{(-1)}\hat {g}_{(0)}^{-1}.
\end{equation}
\subsection{W-gauges}
In accordance with \cite{4}-\cite{7}, the WZNW dynamics is best
seen by going to other gauges which we characterize
with a superscript $(W\pm )$. They are gauges
of the axial type where one component,
that is,  $A_\mp ^{(W\pm)}$ vanishes. In this subsection,
 we rederive results
of refs.\cite{4}-\cite{7}, for completeness.
The change of gauge is most easily achieved starting
from the T-gauge. Let us write, for the $(W+)$ gauge,
\begin{equation}
A_{\pm}^{(W+)}=\omega _{(+)}A_{\pm}^{(T)}\omega _{(+)}^{-1} +
\omega _{(+)}\partial _{\pm}\omega_{(+)}^{-1}, \>
\hbox {that is}, \> g^{(W+)}=g^{(T)}\omega_{(+)}^{-1}.
\label{W1}
\end{equation}
It follows from Eq.\ref{15} that, if
\begin{equation}
\omega _{(+)}^{-1}\partial _+
\omega _{(+)}=\hat {g}_{(0)}^{-1}X_{(1)} \hat {g}_{(0)},
\label{W2}
\end{equation}
then
\begin{equation}
A_{+}^{(W+)}= 0,\qquad A_{-}^{(W+)}\equiv -J=
\omega_{(+)} ( X_{(-1)} + \hat {g}_{(0)}^{-1} \partial_-
\hat {g}_{(0)} +\partial_-) \omega_{(+)}^{-1}.
\label{W23}
\end{equation}
It is clear from Eq.\ref{W2} that $\omega_{(+)} \in G_+$ as
the notation anticipated. Thus the W-gauges are
 not triangular. On the other hand, it is easy to see that
 $J$ satisfies
\begin{equation}
\partial_+ J=0,\quad J_{(-)}=-X_{(-1)}.
\label{W3}
\end{equation}
The first equation follows from
 the zero-curvature condition Eq.\ref{2},
since $A_{+}^{(W+)}= 0$. The second may be verified from the
explicit expression
just given  (following the general convention,  $J_{(-)}$ is
the negative-gradation component of $J$).

The other W gauge is treated similarly, starting
from the other Toda gauge potentials $\tilde A_{\pm}^{(T)}$:
\begin{equation}
A_{\pm}^{(W-)}=\omega_{(-)}^{-1}\tilde A_{\pm}^{(T)}\omega _{(-)} +
\omega _{(-)}^{-1}\partial _{\pm}\omega_{(-)}, \>
\hbox {that is}, \> g^{(W-)}=\tilde g^T\omega_{(-)};
\label{W4}
\end{equation}
\begin{equation}
\omega _{(-)}\partial _+
\omega _{(-)}^{-1}=\hat {g}_{(0)}X_{(-1)} \hat {g}_{(0)}^{-1};
\label{W5}
\end{equation}
\begin{equation}
A_{-}^{(W-)}= 0,\qquad A_{+}^{(W-)}\equiv -\bar J=
\omega_{(-)}^{-1} ( X_{(1)} + \hat {g}_{(0)} \partial_-
\hat {g}_{(0)}^{-1} +\partial_+) \omega_{(+)}^{-1};
\label{W6}
\end{equation}
\begin{equation}
\bar J_{(+)}=X_{(1)},\quad  \partial_-\bar J=0.
\label{W7}
\end{equation}
In accordance with \cite{4}-\cite{10}, the group element
$\omega$ which
parametrises  the solution of the constrained
WZNW equation is defined as follows:
\begin{equation}
\omega\equiv \omega=\omega _+\, \omega _0\,\omega _-;
\> \hbox{where}\> \omega _0=\hat {g}_{(0)}^{-1}.
\label{21}
\end{equation}
Indeed, by explicit computation one verifies that the
currents $J$ and $\bar J$  introduced above are the
corresponding WZNW currents, that is,
\begin{equation}
J\equiv \partial _-\omega\,\omega ^{-1},\;
\bar J\equiv -\omega ^{-1}\partial _+\omega.
\label{23}
\end{equation}
Therefore, Eqs.\ref{W3}, and \ref{W7}
show that $\omega$ is a solution of the
corresponding conformally reduced WZNW model.
\section{Explict solution of the  WZNW model}
As noted in refs. \cite{4} - \cite{7}, the Toda field
$\hat g_{(0)}$ is recovered from  $\omega$ by using the fact that
Eq.\ref{21} is a Gauss decomposition, so that, when we compute
matrix elements between states $| \lambda_j >$
which are annihilated
by ${\cal G}_+$, we get $< \lambda_j| \omega | \lambda_j >
=< \lambda_j|\hat g^{-1}_{(0)} | \lambda_j >$. Of course $\omega$
contains more informations. It is the purpose of the present section
to establish its full
 connection with the group-elements which appeared
in the triangular gauges of subsections 3.1 and 3.2.
 It follows from
Eqs.\ref{W2}, and \ref{W5}  that
\[
\omega _{(+)}^{-1}\partial _+\omega _{(+)}
=(g_{(0)}\, y_{(0)-})^{-1}\, L_{(1)}\,
g_{(0)}\,
y_{(0)-}
=y_{(0)-}^{-1}\, N_{(1)}^{-1}\partial _+N_{(1)}\, y_{(0)-};
\]
\[
\omega _{(-)}\partial _-\omega _{(-)}^{-1}=
\left (g_{(0)}^{-1}\, y_{(0)+}\right )^{-1}
\, L_{(-1)}\, g_{(0)}^{-1}
y_{(0)+}=y_{(0)+}^{-1}\, N_{(-1)}^{-1}\, \partial _-N_{(-1)}\,
y_{(0)+}.
\]
Since $\omega _{(\pm )}\in {\cal G}_{\pm}$, a more convenient form
of these formulae is
\[
\omega _{(+)}^{-1}\partial _+\omega _{(+)}=
{}\left [y_{(0)-}\, N_{(1)}\, y_{(0)-}^{-1}\right ]^{-1} \partial _
+\left [y_{(0)-}\,  N_{(1)}\,  y_{(0)-}^{-1}\right ];
\]
and
\[
\omega _{(-)}\partial _-\omega _{(-)}^{-1}=
{}\left [y_{(0)+}^{-1}\, N^{-1}_{(-1)}\, y_{(0)+}\right ]
\partial _-
\left [y_{(0)+}^{-1}\, N^{-1}_{(-1)}\, y_{(0)+}\right ]^{-1}.
\]
where we used the fact that $\partial_\mp y_{(0)\pm}=0$. Thereof,
\begin{equation}
\omega _{(+)}
=q_{(+)}(z_-)\, y_{(0)-}\, N_{(1)}\, y_{(0)-}^{-1},\quad
\omega _{(-)}=
y_{(0)+}^{-1}\, N_{(1)}^{-1}\,  y_{(0)+} \, q_{(-)}(z_+),
\label{25}
\end{equation}
where $ q_{(+)}(z_-)$ and $ q_{(-)}(z_+)$ are arbitrary elements  of
$ G_+$ and $ G_-$, respectively.
Consequently,
\begin{eqnarray}
\omega & = & q_{(+)}(z_-)\, y_{(0)-}\,  N_{(1)}\> g_{(0)}^{-1}\>
N_{(-1)}^{-1}\, y_{(0)+}\, q_{(-)}(z_+)
\nonumber\\
 & = & q_{(+)}(z_-)\, y_{(0)-}\,  M_{(-1)}^{-1}\> M_{(1)}\,
 y_{(0)+}\,  q_{(-)}(z_+).
\label{26}
\end{eqnarray}
Here $ q_+(z_-)$ and $ q_-(z_+)$ are arbitrary elements  of
$ G_+$ and $ G_-$, respectively.
Since $y_{(0)\pm}$ and $M_{(\pm 1)}$ are only functions
of $z_\pm$ respectively, this shows  that
  $\omega ^{-1}$ has the holomorphic decomposition
\begin{equation}
\omega ^{-1} = g_L(z_+)\,g_R(z_-),
\label{27}
\end{equation}
\begin{equation}
g_L(z_+)\equiv q_{(-)}^{-1}(z_+)\,h^{-1}_{(0)+}(z_+)\,
M_{(1)}^{-1}(z_+),\,
g_R(z_-)\equiv M_{(-1)}(z_-)  h^{-1}_{(0)-}(z_-)\,  q_{(+)}^{-1}(z_-).
\label{28}
\end{equation}
This of course agrees  with the fact that $\omega$
is a solution of the
WZNW equations $\partial_+ J=\partial_-\bar J =0$.
As is well known\cite{3},  $g_R$ and $g_L$ are solutions of the
equations $d g_R /dz^-=-g_R J$, $ d g_L /dz^+ = \bar J g_L$.
These holomorphic relations have been called generalised Bargmann,
or generalised Frobenius,  or Drinfeld-Sokolov equations.
The arbitrary functions $q_{(\pm)}^{-1}(z_\mp)$ reflect the usual
gauge invariance of the WZNW solution, and of the holomorphic
equations just recalled.

The elements $g_{L,R}$ can be written in the form of the modified
Gauss decomposition, namely
\begin{equation}
g_{L,R}\,=\,g_{(-)L,R}\,g_{(0) L,R}\,g_{(+)L,R};
\label{29}
\end{equation}
where
\begin{eqnarray}
g_{(-)L} & = & q_{(-)}^{-1},\;g_{(0)L}=y_{(0)+}^{-1},\;
 g_{(+)L}=M_{(1)}^{-1};
\nonumber\\
g_{(-)R} & = & M_{(-1)},\;g_{(0)R}=
y_{(0)-}^{-1}, \;g_{(+)R}=q_{(+)}^{-1}.
\label{30}
\end{eqnarray}
Finally, we write down a few useful relations,
\[
\partial _+g_{(+)L}\equiv \partial _+M_{(1)}^{-1}=
-L_{(1)}M_{(1)}^{-1}=
-L_{(1)}g_{(+)L}\equiv -\bar {F}(z_+)g_{(+)L},
\]
and
\[
\partial _-g_{(-)R}\equiv \partial _-M_{(-1)}=M_{(-1)}L_{(-1)}=
g_{(-)R}L_{(-1)}\equiv g_{(-)R} {F}(z_-).
\]
Thereof,
\begin{eqnarray}
\bar {F}(z_+) & = & L_+=\sum _{\alpha}\phi_{+\alpha}X_{(1)\alpha}
=h_{(0)+}X_{(1)}(h_{(0)+})^{-1}=
g_{(0)L}^{-1}X_+g_{(0)L},\nonumber\\
F(z_-) & = & L_-=\sum _{\alpha}\phi_{-\alpha}X_{(-1)\alpha}=
h_{(0)-}X_-(h_{(0)-})^{-1}=g_{(0)R}X_{(-1)}g_{(0)R}^{-1}.
\label{31}
\end{eqnarray}

The explicit form of the solution of conformally reduced
WZNW equation is crucial for the geometrical interpretation
of the corresponding Toda systems. Indeed, as shown in
ref. \cite{9}, the solutions of the
holomorphic equations
$d g_R /dz^-=-g_R J$, $ d g_L /dz^+ = \bar J g_L$,   for $A_n$-Toda
in a certain gauge, coincide with the embedding functions that
defines the associated W-holomorphic surface in $CP^n$.
This result may be generalized to any conformal Toda
theory\cite{17}.

\section{Outlook}

Besides the W-geometrical  aspect just mentioned,
the transition between the matrix elements and
the connections given
in different gauges is of a great importance for
the study of various aspects
of the Toda-type theories, in particular for a
formulation of the boundary value problem in terms of
the characteristic
integrals for the system (\ref{T}).

One may, of course, also consider the case of half-integral
gradations. These can be incorporated into the scheme by
extending $A_{\pm}$ so that they take their values in
${\cal G}_{\pm{1
\over 2}}$ as
well as ${\cal G}_0$ and ${\cal G}_{\pm 1}$. If
	there are no further
constraints, the field equations (\ref{T}) then
generalize to a set \cite{3,12}
\begin{eqnarray}
&\partial_ + (\hat g_{(0)}^{-1}
\partial_- \hat g_{(0)}) = [X_-,\, \hat g_{(0)}^{-1}
X_+ \hat g_{(0)}] +
[\widetilde \psi,\, \hat g_{(0)}^{-1}\psi \hat g_{(0)}],&
\nonumber \\[-0.5em]
\label{4'}
\\[-0.5em]
&\partial_- \psi = [X_{(1)},\, \hat g_{(0)}
\widetilde \psi \hat g_{(0)}^{-1}],
\qquad
\partial_ + \widetilde \psi = [X_{(-1)},\, \hat g_{(0)}^{-1} \psi
\hat g_{(0)}];&\nonumber
\end{eqnarray}
and are derivable from  the effective action \cite{7}
\begin{eqnarray}
I^{\mbox{ eff }} & = & I(\hat g_{(0)})
-\int \hbox{tr}\bigl( X_{(1)} \hat g_{(0)}^{-1} X_{(-1)}
\hat g_{(0)}\bigr) +
\int \hbox{tr}\bigl( (\partial_+ \psi) \hat g_{(0)} (\partial_-
\tilde \psi)\hat
g_{(0)}^{-1} \bigr)  \nonumber \\
& + &
\int \hbox{tr}([X_{(-1)},\psi]\partial_+ \psi)+
 \int \hbox{tr}([X_{(1)},\tilde \psi]\partial_-  \tilde \psi),
\end{eqnarray}
where $\psi$ and $\tilde \psi$ are the grade $\pm{1 \over 2}$ fields.
On the other hand,  one may start from the principle
that the constraints
 be a maximal set of first-class
constraints, as in \cite{7}; then
half of the grade-$({1 \over 2})$ fields $\psi$ must be zero, and
the effective action takes the more complicated form
\begin{eqnarray}
I^{\mbox{ eff }} & = & I(\hat g_{(0)}) -\int \hbox{tr}
\bigl( X_{(1)} \hat g_{(0)}^{-1} X_{(-1)}
\hat g_{(0)}\bigr)\nonumber \\
& - & \int < (\partial_+ \eta) (A-BD^{-1}C)(\partial_- \tilde \eta)>
 -\int <[X_{(-1)},\partial_+ \eta], BD^{-1}\eta>\nonumber \\
& + & \int <\tilde \eta D^{-1}C,
[X_{(1)},\partial_-
\tilde \eta ]> - \int <[X_{(-1)},\eta ] D^{-1} [X_{(1)},\tilde
\eta ]>,
\end{eqnarray}
where the $\eta$'s are the remaining grade-$(\pm{1 \over 2})$
fields, $<,>$ denotes the restriction of the Cartan inner-product
to these fields and the matrices $\{A,B,C,D\}$ are defined as
$\hbox{Ad}g_0=\pmatrix{
A & B  \cr
C & D  \cr   }$. In fact,
this principle, has been implicitly
applied in the case of the integral
gradations allowing to rid off $\mbox{ dim }{\cal G}_0 - \mbox{ dim }
{\cal G}_{\pm 1}$ additional  (nonprimary) fields in the stringy
non-abelian systems
associated with non-principal $sl(2, {\bf C})$ subalgebras of $B_r$
\cite{1}.
This  turned out
to be especially important for the study of  non-trivial
background
metrics  in the target space, e.g., for the black holes generated by
the
non-abelian $B_2$-Toda fields, and their $osp(2|4)$ superextension,
see
\cite{10,13}.
It is interesting to mention
 related reasonings
of the article  \cite{14} which  argue that the
weight-$1/2$ fermion fields
in the superconformal field theory
are ``auxiliary'' ones, and that their
role is only  to ensure  the closure of the corresponding
Lie superalgebra.  A
similar conclusion takes place for the aforementioned
non-abelian $B_2$ (and
$osp(2|4)$)  -- Toda systems, as well
as for their generalizations to  the Lie
(super)algebras of higher ranks.

Consider  the supersymmetrical extensions\cite{15}
of the equation (\ref{T}),
\begin{equation}
{\cal D}_+\left ( \hat g_{(0)}^{-1}{\cal D}_-  \hat g_{(0)} \right )=
\left [ Y_{(-1)} ,\,  \hat  g_{(0)}^{-1} Y_{(1)}
\hat g_{(0)}\right ]_+ ,
\label{5}
\end{equation}
associated with a classical Lie algebras and
 superalgebra ${\cal G}^s$ supplied
with a
${\bf Z}$-gradation. Here  ${\cal D}_{\pm }$ are the
supercovariant derivatives
in $2|2$-superspace; $\hat g_{(0)}$ is a regular
element
of the Grassmann span of the Lie group $G_0=\mbox { Lie }{\cal G}_0$;
$Y_{(\pm 1) }$ are fixed (odd)
elements of ${\cal G}_{\pm 1}$. Most of the
results given above  can be
derived  for the system (\ref{5}) as well, with
the corresponding  modifications. (For the relation with a
supersymmetric extension of the constrained-WZNW-model
see also \cite{16}.)
It is interesting to emphasize that,  under the relevant conditions,
the
component-form of Eqs.\ref{5}  coincides with Eqs.\ref{4'}.
This is, of course,
 only formal, since the fermionic components entering in Eq.\ref{5}
are
Majorana spinors with anticommuting values,
contrary to  the functions $\psi $ and $\tilde \psi $ in
Eq.\ref{4'}.

Finishing up this paper, we would like to mention that
most of the relations
and conclusions given here are applicable to the
periodic Toda-type system
associated with the affine Lie Kac-Moody algebras
$\tilde {\cal G}$,
of course, with the appropriate modifications. For
many interesting cases, these equations, in particular those
called last time also as affine and conformal affine Toda
theories, ---  which are
given in  quite general a form in ref.\cite{3}
following the results of ref.\cite{18},  together
with their group-algebraic
integration scheme ---
 can be rewritten  in the form Eq.\ref{T}.
Here the meaning of
the elements $X_{(\pm 1)}$ is different, however,
and the relevent
grading operator ${\cal H},
[{\cal H}, {\cal G}_m]=m{\cal G}_m$, of course
does not belong
to the corresponding
finite-dimensional Lie algebra ${\cal G}$. It can be
constructed in a degenerate
representation of $\tilde {\cal G}$ with the
spectral parameter
$\lambda $ as ${\cal H}=H+c\,\lambda d/d\lambda $.
Then, following the same reasonings as in
ref.\cite{10} for the construction of a nontrivial
background metric in the target
space, one can describe
the black holes associated with the
non-canonical gradation of the affine
algebra $B_2^{(1)}$, and  the
soliton solutions of the model, for example.
Here, at  a
formal level, the difference between
the corresponding finite and  affine cases
arises from   the insertion,  in the
expression for $X_{(\pm 1)}$,  of the additional
term like $\lambda ^{\pm 1} X_{\mp M}$,
where $M$ is the
 maximal root of $B_2$. (In the general case a
modification of $X_{(\pm 1)}$ is the relevant
element of the Heisenberg subalgebra of the
corresponding affine Lie algebra.) The final
equations, of course, do not
depend on $\lambda $, and provide the simplest
example of a nontrivial background metric,
while the canonical gradation of
the affine Lie algebras leads, as in the
finite-dimensional case, to
constant metrics  in the target space of
the corresponding $\sigma $-model.
It seems quite believable that a
quantization procedure for such
systems, completely integrable at  the
classical level, could follow along
the line of the corresponding construction
given in \cite{3}, and in \cite{19}
 for finite-dimensional
 systems, with the necessary modification of such objects
as the Casimir operator of the second order, the Whittaker
vectors, etc.

\bigskip

{\bf Acknowledgements.}  One of the authors (M.S.) would like to
thank LPTENS in Paris, ENSLAPP in Lyon, IAS in Dublin, and
the Isaac Newton
Institute in Cambridge for kind hospitality.
His research was supported in part
by SERC Visiting fellowship, grant GB.929.612/RG.14173-MTA-R3.
This work  was partially supported by the European  Twinning
Program, contract   \#  540022.


\bigskip

\begin{thebibliography}{99}
\bibitem{1}
A.~N.~Leznov, M.~V.~Saveliev: Phys. Lett. B, v.~79 (1978) 294;
Lett. Math. Phys. v.~3 (1979) 207; Comm. Math.
Phys. v.~74 (1980) 111.
\bibitem{2}
A.~N.~Leznov, M.~V.~Saveliev:
Lett. Math. Phys. v.~6 (1982) 505; Comm. Math.
Phys. v.~89 (1983) 59.
\bibitem{3}
A.~N.~Leznov, M.~V.~Saveliev:
{\it Group-Theoretical Methods for Integration
of Nonlinear Dynamical Systems.} Progress
in Physics v.~15, Birkhauser-Verlag,
1992.
\bibitem{G}  J.-L. Gervais,  Comm. Math. Phys.
 130 (1990) 257; 138 (1991) 301.
\bibitem{4}
J.~Balog, L.~Feh\'er, L.~O'Raifeartaigh, P.~Forg\'acs, A.~Wipf:
Ann. Phys. (N.Y.), v.~ 203 (1990) 76.
\bibitem{5}
L.~O'Raifeartaigh, P.~Ruelle, I.~Tsutsui, A.~Wipf: Comm. Math. Phys.
v.~143 (1992) 333.
\bibitem{6}
L.~Feh\'er, L.~O'Raifeartaigh, P.~Ruelle,
I.~Tsutsui, A.~Wipf: Ann, Phys.
(N.Y.), v.~213 (1992) 1.
\bibitem{7}
L.~Feh\'er, L.~O'Raifeartaigh, P.~Ruelle,
I.~Tsutsui, A.~Wipf: to appear in
Phys. Reports.
\bibitem{8}
J.-L.~Gervais, Y.~Matsuo: Phys. Lett. B, v.~274 (1992) 309;
\bibitem{9}
J.-L.~Gervais, Y.~Matsuo: {\it $W$-Geometries.}
Preprint LPTENS-91/29, 1992;
to appear in  Comm. Math. Phys.
\bibitem{10}
J.-L.~Gervais, M.~V.~Saveliev: Phys. Lett. B, v.~286 (1992) 271.
\bibitem{11}
S.~Cecotti, C.~Vafa: Nucl. Phys. B, v.~367 (1991) 359.
\bibitem{12}
A.~N.~Leznov: {\it Exactly Integrable
Systems with Fermionic Fields.} Preprint
IHEP 83-7, Serpukhov, 1983.
\bibitem{13}
F.~Delduc, J.-L.~Gervais, M.~V.~Saveliev:
Phys. Lett. B, v. 292 (1992) 295.
\bibitem{14}
P.~Goddard, A. ~Schwimmer: Phys. Lett. B, v.~214 (1988) 209.
\bibitem{15}
M.~V.~Saveliev: Comm. Math. Phys. v.~95 (1984) 199;\\
A.~N.~Leznov, M.~V.~Saveliev:
Sov. J. Theor. Math. Phys., v.~61 (1984) 150;\\
D.~A.~Leites, M.~V.~Saveliev,
V.~V.~Serganova: -- in {\it Group Theoretical
Methods in Physics.} V.~I.~Manko, M.~A.~Markov, eds., VNU Sci.
Press, v.~1 (1986) 255.
\bibitem{16}
F.~Delduc, E.~Ragoucy, P.~Sorba: Comm. Math. Phys. v.~146 (1992) 403.

\bibitem{17}  J.-L.~Gervais, M.~V.~Saveliev, in preparation.

\bibitem{18}
A. N. Leznov, M. V. Saveliev, V. G. Smirnov: Sov. J. Theor.
Math. Phys., v. 48 (1981) 3.

\bibitem{19} A.Bilal, J.-L.Gervais:
Phys. Lett. { 206B} (1988), 412; Nucl. Phys.
{ 314B} (1989) 646; {318B} (1989) 579.


\end{thebibliography}
\end{document}